# Looking through walls and around corners with incoherent light: Wide-field real-time imaging through scattering media

Ori Katz[*,†], Eran Small[†], and Yaron Silberberg

Department of Physics of Complex Systems, Weizmann Institute of Science, Israel

ori.katz@weizmann.ac.il

[†]authors contributed equally to this work

**Imaging with optical resolution through highly scattering media is a long sought-after goal with important applications in deep tissue imaging. Although being the focus of numerous works[1], this goal was considered impractical until recently. Adaptive-optics techniques[2-4] which are effective for correcting weak wavefront aberrations, were deemed inadequate for turbid samples, where complex speckle patterns arise and light is scattered to a large number of modes that greatly exceeds the number of degrees of control[5]. This conception changed after the demonstration of focusing coherent light through turbid media by wavefront-shaping, using a spatial-light-modulator (SLM)[6,7]. Here we show that wavefront-shaping enables widefield real-time imaging through scattering media with both coherent and incoherent illumination, in transmission and reflection. In contrast to the recently introduced schemes for imaging through turbid media[8-14], our technique does not require coherent sources[8-14], interferometric detection[10-14], raster scanning[8-10,14] or off-line image reconstruction[11-14]. Our results bring wavefront-shaping closer to practical applications, and realize the vision of looking 'through walls' and 'around corners'[15].**

The ability to image through inhomogeneous media is extremely valuable in numerous applications, ranging from astronomic observation through the turbulent atmosphere to microscopic imaging in turbid tissues. Between these extremes exist various mundane tasks such as looking at foggy scenes or peeking through shower curtains. The reason that inhomogeneous media limit optical observation is that the random refractive index variations distort the spherical wavefronts generated by every point on the imaged object; resulting in a smeared (and sometimes speckly[5]) image (Fig.1a-b).

Here we present a technique based on high-resolution wavefront shaping[6,7] that enables direct real-time wide-field imaging in three-dimensions of structures hidden behind highlyscattering and turbid media. Our technique exploits the angular range where a single wavefront correction holds, which is dictated in multiple-scattering media by the optical 'memory effect' for speckle correlations[16,17]. Remarkably, although wavefront-shaping is based on forcing constructive interference, our technique works for both coherent and incoherent illumination. The ability to image incoherently illuminated objects is an advantage, as it eliminates speckle imaging artifacts and is well-matched with the common light sources used in microscopy. Wavefront shaping has been recently shown to enable imaging with coherent illumination through turbid, nearly opaque samples[8-14]. This was achieved either by separately measuring the correction for every point on the object[11-14], or by raster-scanning the wavefront-shaped focus in the limited range dictated by the memory-effect[8-10]. Our technique departs from these schemes in the use of incoherent light and in the fact that after a single optimization procedure, the object's image is formed in real-time on a standard camera (Fig.1c), without the need for slow raster scanning[8-10] or computerized image reconstruction[11-14].

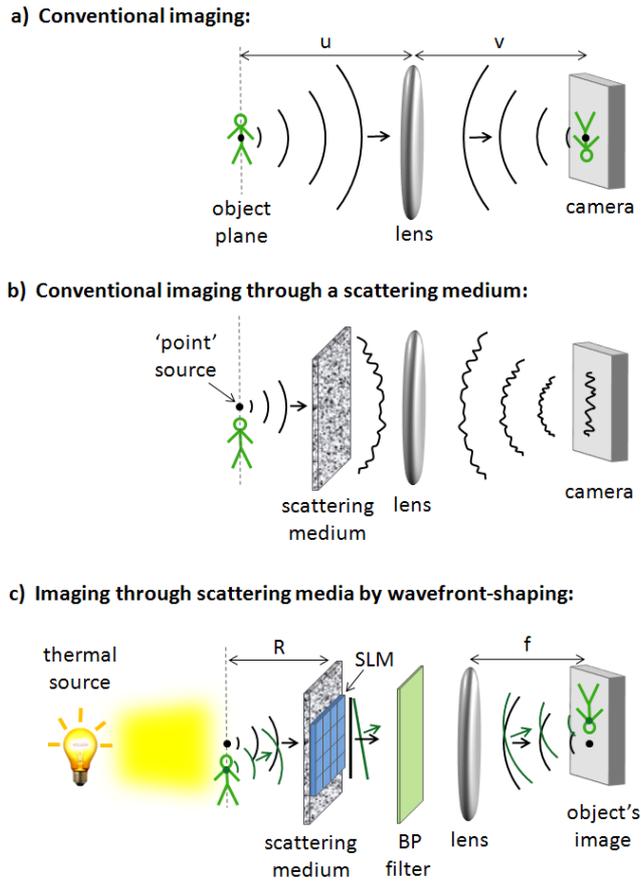

**Figure 1. Schematic of the experiment: a,** in conventional imaging, a lens is used to create an image of an object on a camera plane, by refocusing the light from any point source on the object plane; **b,** when a highly scattering medium blocks direct view of the object, the light from any point source on the object plane is scattered, resulting in a diffuse light field and random speckle patterns on the camera. **c,** In the presented technique, a spatial light modulator (SLM) shapes the phase of the scattered light from a point-source placed at a distance R behind the medium, such that it interferes constructively at a desired point on the camera and re-focuses[6]. The SLM thus effectively transforms the medium into a "scattering lens" with a focal length R. As result of the memory-effect[16, 17] the single wavefront correction is effective for nearby points, and any object placed in the vicinity of the point-source position is imaged directly on the camera in real-time. A narrow bandpass filter rejects uncorrected wavelengths[18]. Note that the lens in (c) is not obligatory, since the scattering medium/SLM combination can generate directly the converging wave required for imaging. (4-f telescope for imaging the SLM on the medium not shown).

A schematic of the experiment is presented in Fig.1c and representative results are given in Fig.2. A point-source is placed at some distance R behind the scattering medium. On the other side of the medium, the scattering medium's surface is 4-f imaged to the plane of a two-dimensional phase-only SLM (4-f telescope not shown in Fig.1c, see Supplementary Material). The phase-shaped light is then bandpass filtered and focused on a CCD camera by a lens with a focal length f. For a flat-phased SLM, the point source produces a random speckle pattern on the camera (Figs. 1b, 2a). However, the intensity of one of the speckles can be easily enhanced 1000-fold by displaying the appropriate SLM phase-pattern, which can be found by adaptive optimization[6]. As a result, a bright spot with a contrast proportional to the number of controlled SLM segments, $N_{SLM}$, appears on the camera[6] (Fig. 2b). This intensity enhanced speckle is the corrected image of the point source (Fig.1c). As result of the memory-effect[24, 25] shifting the point-source in the transverse direction will result in addition of a linear phase ramp (tilt) to the transmitted wavefront, which will be transformed by the lens to a correlated shift of the point-source image on the camera. Furthermore, for the same reason, any object placed in the vicinity of the pre-corrected point-source will be directly imaged on the camera, without requiring any additional optimization or calculation (Fig.1c). The results obtained with an object illuminated with an incoherent white-light thermal source (Tungsten-halogen lamp, Fig.2d-e) and a coherent monochromatic plane-wave (Fig.2f), demonstrate the elimination of speckle artifact with incoherent illumination.

The imaging mechanism can be understood as follows: Essentially, after the optimization procedure with a point-source, the optimized SLM phase-pattern transforms the scattering medium into a 'scattering-lens' with a focal length R at the corrected wavelength[6, 9, 15]. The entire system (scattering-medium - SLM - lens) is then a telescopic imaging system with a magnification of M=f/R, i.e. a widefield 'scattering microscope' (Figs.1c). The field of view (FOV) of this 'scattering microscope' is limited by the angular range for which the single wavefront correction holds. This is the range in which translating the point-source produces correlated shifted speckle patterns on the camera[15, 17]. For diffuse light propagation through a multiple-scattering medium of thickness L, this range is given by the memory-effect[15, 16]: FOV≈R$\lambda$/$\pi$L. For a thin scattering surface the FOV is in theory unlimited, but in practice will be limited by the accuracy by which the SLM and the medium planes are conjugated in the axial position, $\varepsilon$: FOV~R$\sigma_x$/$\varepsilon$, where $\sigma_x$ is the transverse correlation length of the shaped/scattered field (See Supplementary Material). For the case of a thick scattering medium that does not completely diffuse the light, the FOV is related to the isoplanatic patch size[19]. An additional limit for the FOV is the finite aperture of the scattering lens, D, which will cause vignetting. For a thin scattering medium, D is the diameter of the SLM image on the medium surface. The memory-effect can also be used to derive the range by which the object can be translated *axially* and still be imaged on the properly translated camera plane. In the paraxial limit, this range is approximately $\Delta z_{FOV}$≈2FOV·R/D (see Supplementary Material). To summarize, after displaying the correct phase-pattern, one can image an object located in the three-dimensional volume of FOV×FOV×$\Delta z_{FOV}$ around the corrected point-source.

The transverse resolution of the scattering-microscope at the object plane, $\delta$x, is dictated by the numerical-aperture of the scattering lens, and is given by[7, 9]: $\delta$x=1.03$\lambda$R/nD, where n is the refractive index behind the medium. In a high index medium, such lens can provide a resolution better than 100nm[9]. Similarly, the axial resolution (depth of focus) is given by[15]: $\delta z = (2\lambda/\pi)(R/D)^2$.

The presented scheme can also work in reflection, i.e. image occluded objects using the diffused back-scattered light reflected from a random medium (e.g. a wall). In such a scenario (Fig. 3a), the optimized phase-pattern effectively transforms the scattering medium into a curved 'scattering mirror' at the corrected wavelength. Thus realizing the vision of 'looking around corners' using scattered light[15]. In Figure 3 we demonstrate the ability to image in reflection by imaging and tracking in real-time incoherently illuminated objects using the diffused back-scattered light from a standard white paper. The imaging resolution and FOV in reflection geometry are given by the same expressions as for transmission geometry, but by replacing the medium thickness, L, with the scattering mean-free-path l*[15]. The stronger the scattering (shorter mean-free-path, and light penetration depth) the larger is the FOV. Interestingly, light at any incident angle can be used for imaging. The optimized phase-pattern can be found for any point-source position, allowing 'tilting' of the 'scattering mirror' and imaging at any angle without changing the reflection collection angle.

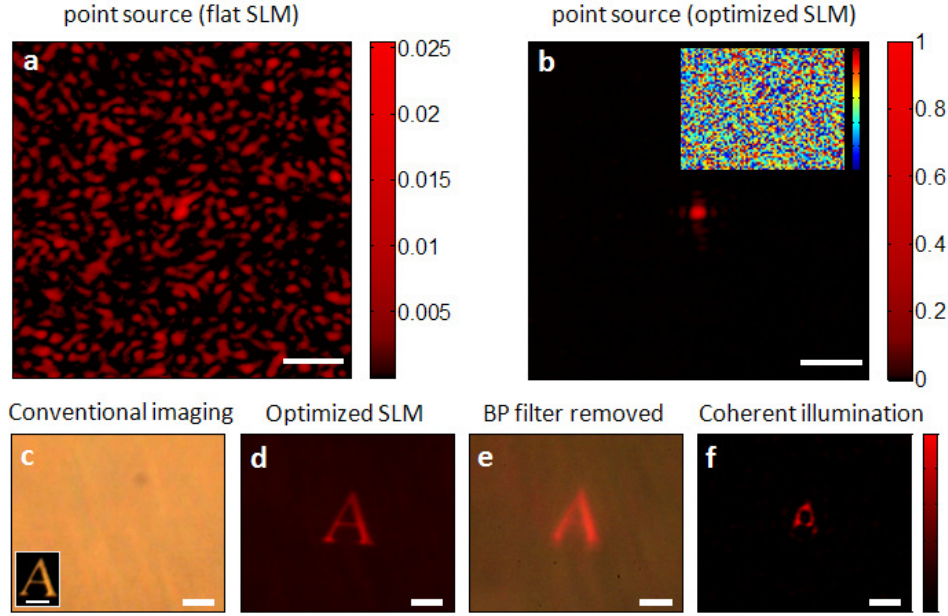

**Figure 2. Widefield imaging through a scattering-medium, experimental results. (a-b)** images of a point-source (50μm pinhole) placed 36.5cm behind the scattering medium: **a,** without correction (flat SLM phase pattern). **b,** after optimizing the SLM phase-pattern to enhance the intensity of a single speckle[6]. The obtained intensity enhancement is 800 (inset: SLM phase-pattern, maximum applied phase is 3π). **(c-e)** Imaging a transmission object (the letter 'A') illuminated by a white-light thermal source (Tungsten-halogen lamp): **c,** conventional imaging through the scattering medium with a color camera. Inset: direct imaging without the scattering medium; **d,** imaging through the scattering medium using the optimized phase-pattern of (b). **e,** same as d but with the bandpass filter removed (taken with a color camera). The contrast is reduced due to uncorrected spectral components[18]; **f,** imaging through the scattering medium using coherent illumination (collimated HeNe laser beam) and a 40% smaller object. Speckle artifacts are apparent. Scale-bars, 3mm.

The two fundamental limitations of our scheme are its frequency-bandwidth and image contrast. The frequency-bandwidth of the scattering-microscope is the same as that of the focus formed by wavefront shaping[6], and is given by the speckle spectral correlation width, $\Delta f$[18]. The optimized phase-pattern may thus be used to form an image only for spectral components within $\Delta f$. Any other uncorrected spectral components will result in a reduced image contrast (Fig.2e). The narrow frequency bandwidth is one of the distinguishing characteristics of wavefront-shaping[6-13] compared to adaptive-optics. The reason for the limited bandwidth is that the wavefront-shaped focus is formed by interfering many different optical paths[18], whereas in adaptive-optics[2,3] all the optical paths have the same length. The results of Fig.2d-e illustrate that the key for the imaging ability is the use of spectral components only within this limited frequency-bandwidth.

The image contrast for a single point source object is the intensity enhancement of a single speckle relative to the average background[6]: $\eta_0 \approx (\pi/4) N_{SLM}$. Different from adaptive optics, a sharp image with a high contrast of $\sim N_{SLM}$ is obtained even when the number of scattered modes collected by the scattering-lens (number of speckles) is much larger than the number of degrees of control, $N_{SLM}$ (Fig. 2b). Most importantly, the imperfect correction is manifested only in reduced intensity and contrast of the focus and not in loss of resolution[6,7,9]. For a general incoherently illuminated object that is composed of N bright resolution cells, the contrast will be reduced to: $C \approx \eta_0/N$, where N is given approximately by the bright area of the object, A, divided by the area of a single resolution cell: $N \approx A/\delta x^2$. This unconventional trade-off between imaging resolution ($\delta x$) and contrast can be understood from the fact that the point-spread-function of the scattering-microscope is a bright peak on top of a speckled background with a peak to background ratio of $\eta_0$ (Fig. 2b).

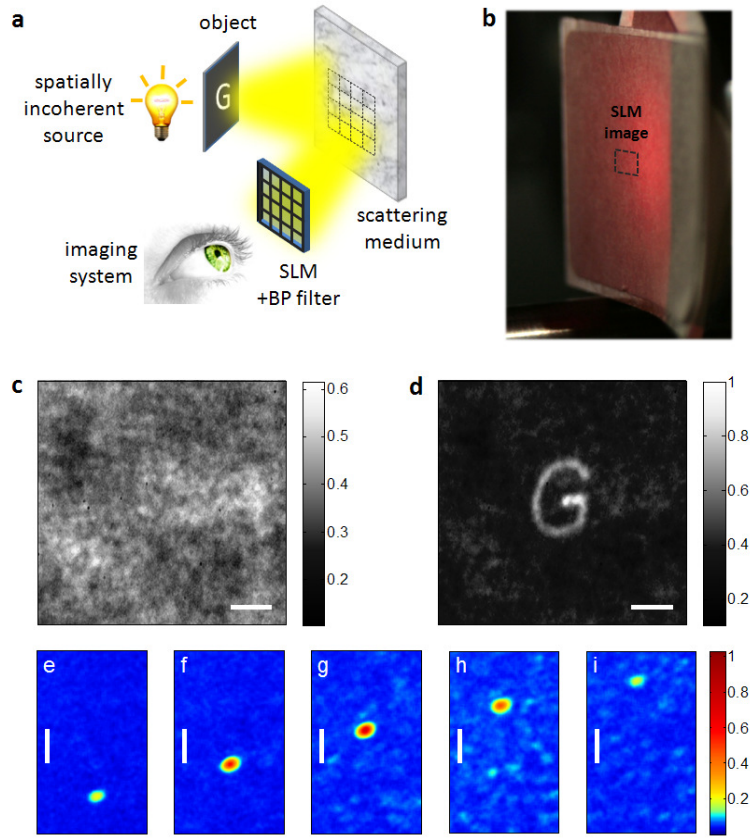

**Figure 3. Looking 'around corners' with incoherent light. a,** An SLM is 4-f imaged on the surface of a highly-scattering medium (in this example, a white paper). The SLM is used to correct the wavefront distortions of the diffused back-scattered light. After performing the correction for a point-source, the reflecting medium is transformed into a 'scattering mirror', allowing imaging in real-time of objects placed in the vicinity of the pre-corrected point-source. **(b-i)** experimental results using the light reflected from a piece of white paper: **b,** the back-scattered light as seen by the naked eye. **c,** image of the object (transmission plate with the letter 'G', placed 30cm from the scattering medium) before correction (flat SLM phase-pattern), and **d,** image obtained with the optimized phase-pattern. **e-f,** Real-time tracking of a moving occluded object (240μm pinhole at 30cm distance) using the light scattered from the paper (see Supplementary Movie). Scale-bars, 1mm.

The use of spatial wavefront-shaping for controlling light in scattering media has fascinating analogues in time-reversal techniques in acoustics and radiofrequency electromagnetic waves[20]. Related notions for imaging through random media in optics can be traced back to phase-conjugation/phase-subtraction holographic techniques[21-25]. These, however, require coherent laser sources and interferometric detection. The ability to use incoherent light for widefield imaging is one of the major advantages of our technique in applications such as microscopy.

In our experiments we focused on situations where the measured light intensity is high, allowing direct imaging using incoherent sources and standard CCD cameras. Using EMCCD cameras will shorten the acquisition times and allow imaging at low light levels, such as those available in transmission through nearly opaque samples[6, 8]. We used a closed-loop feedback algorithm to determine the wavefront correction. Alternatively, transmission-matrix[11, 12, 26] or phase-conjugation[10, 25] approaches may be used. In specific scenarios it may be possible to find the correction without a point-source, using image-based optimization metrics such as contrast. Applying additional control of spatial polarization and amplitude should result in enhanced contrast[6, 27], and exploiting the spatio-temporal degrees of control[28-30] may allow nonlinear imaging using ultrashort pulses and implanted probes[10]. Finally we note that after applying the appropriate phase pattern, the scene behind the medium may be watched with the unaided eye. Unfortunately, the light intensities in our experiments were too low to directly 'see' the scene without a camera.

**Methods**

The complete experimental setups are presented in Supplementary Figures 1-2. All images excluding the color images of Fig.2c,e were obtained with a Watec WAT-120N CCD camera. The color images of Fig.2c,e were obtained with a Cannon EOS-400D SLR camera. Presented images are median-filtered and low-pass filtered with a rectangular 3x3 pixels kernel. In the optimization process, the SLM (Hamamatsu LCOS-SLM X10468) was divided to 4800 (80x60) equally sized square segments, and the phase of the different segments was optimized to maximize the intensity of the light from a point-source at a selected subset of the CCD pixels, using a genetic optimization algorithm. The genetic algorithm was implemented using *Matlab* Genetic Algorithm toolbox, using a population size of 30, 0.4 crossover-fraction, random mutation of 1% of the segments, and elite count of 2. The CCD integration time during the optimization was typically 80ms. For imaging with incoherent illumination the integration times were 2-10s. The scattering samples used are a 10x20° Newport light shaping diffuser (Fig.2), and a white paper sticker with a thickness of 70±10μm (Fig.3). A white-light tungsten-halogen source (Schott KL1500LCD) was used as the light source for imaging in transmission. For imaging in reflection, a narrow bandwidth pseudothermal source was used, as the higher optical losses and the lower correction bandwidth, $\Delta f$, required a more spectrally intense source to achieve integration times of the order of one second. The point-sources were a 50μm pinhole (Fig.2) and a 100μm pinhole (Fig.3) illuminated by a collimated HeNe laser beam. The bandpass filter has a bandwidth of 3.4nm (FWHM) around 633.4nm.

## Acknowledgements

We thank Y. Bromberg for stimulating discussions, and G. Han, Y. Shopen, G. Elazar, B. Sharon, Y. Shimoni and R. Baron for technical assistance. O.K. acknowledges support from the Eshkol Fellowship and E.S. acknowledges support from the Adams Fellowship. This work was also supported by grants from the Israel Science Foundation, ERC grant QUAMI, and the Crown Photonics Center.

## Author contributions

O.K. conceived the idea, O.K. and E.S. designed and performed the experiments and analyzed the data. O.K., E.S. and Y.S. wrote the manuscript.